\documentclass[pra,amssymb]{revtex4}

\usepackage{appendix}
\usepackage{graphicx}
\usepackage{amssymb,amsmath} 
\usepackage{epsfig}
\usepackage[utf8]{inputenc}
\usepackage[T1]{fontenc}
\usepackage{color}
\usepackage{soul}

\def\T{k_B T/U}

\def\B#1{\!\left(#1\right)}
\def\BB#1{\!\left[#1\right]}

\def\pVar{\partial_\xi\!{\rm Var}}
\def\p2Var{\partial^2_\xi{\rm Var}}

\def\bg{\begin{equation}\begin{gathered}}
\def\eg{\end{gathered}\end{equation}}

\def\be{\begin{equation}}
\def\ee{\end{equation}}

\def\bx{\begin{equation}\begin{aligned}}
\def\ex{\end{aligned}\end{equation}}

\def\bee{\begin{equation*}}
\def\eee{\end{equation*}}

\def\la{\langle}
\def\ra{\rangle}

\begin{document}

\title{Locating the quantum critical point of the Bose-Hubbard model through singularities of simple observables}

\author{Mateusz \L\k{a}cki$^{1,2,3}$, Bogdan Damski$^{1}$, and Jakub Zakrzewski$^{1}$} 
\affiliation{\mbox{$^1$ 
Instytut Fizyki imienia Mariana Smoluchowskiego, 
Uniwersytet Jagiello{\'n}ski, 
ulica  {\L}ojasiewicza 11, 30-348 Krak\'ow, Poland}
\mbox{$^{2}$ Institute for Quantum Optics and Quantum Information of the Austrian Academy of Sciences, A-6020 Innsbruck, Austria}
\mbox{$^{3}$ Institute for Theoretical Physics, University of Innsbruck, A-6020 Innsbruck, Austria}		
}

\begin{abstract}
We show that  the critical point of the two-dimensional
Bose-Hubbard model can be easily found through studies of either 
on-site atom number fluctuations or the nearest-neighbor two-point correlation function 
(the expectation value of the tunnelling operator).
Our strategy to locate the critical point is 
based on the observation that the derivatives of these observables with respect 
to the parameter that drives the superfluid-Mott insulator transition 
are singular at the critical point in the thermodynamic limit.
Performing  the quantum Monte Carlo simulations of the two-dimensional Bose-Hubbard model,
we show that this technique leads  to the accurate determination of the position of its critical point.
Our results can be easily extended to the three-dimensional Bose-Hubbard model 
and different  Hubbard-like  models.
They  provide a simple experimentally-relevant way of locating  
critical points in various cold atomic lattice systems. 
\end{abstract}
\maketitle

The amazing  recent progress in cold atom manipulations 
allows for experimental studies of strongly correlated bosonic systems placed in 
lattices of various  dimensions and shapes \cite{Lewenstein,Qsim1,KubaReview}.  The basic physics of such 
systems is captured by the Bose-Hubbard model \cite{FisherPRB1989,JakschPRL1998,Lewenstein,Qsim1,KrutitskyPhysRep2015,KubaReview}, 
whose Hamiltonian has the following deceptively simple form  
\be
\label{H}
\hat{H}=-J\sum_{\la\bf i, j\ra}(\hat{a}_{\bf i}^\dag \hat{a}_{\bf j} 
+ \hat{a}_{\bf j}^\dag \hat{a}_{\bf i})
+\frac{U}{2} \sum_{\bf i}\hat{n}_{\bf i}(\hat{n}_{\bf i}-1), \ \ 
[\hat{a}_{\bf i},\hat{a}_{\bf j}^\dag]=\delta_{\bf ij}, \ [\hat{a}_{\bf i},\hat{a}_{\bf j}]=0, 
\  \hat n_{\bf i}=\hat{a}_{\bf i}^\dag\hat{a}_{\bf i},
\ee
where $\la\bf i,j\ra$ stands for  nearest-neighbor lattice sites $\bf i$ and $\bf j$,
while $\hat a_{\bf i}$ ($\hat a_{\bf i}^\dag$) annihilates (creates) an atom in the ${\bf i}$-th lattice site.
The first term in this Hamiltonian describes nearest-neighbor tunnelling, while
the second one describes repulsive  on-site interactions. Thus, the
former term promotes spreading of atoms across the lattice, which
leads to on-site atom number fluctuations that are being suppressed by the
latter term.

The competition between the tunnelling and interactions  leads to the
quantum phase transition
when the average number of
atoms per lattice site (the filling factor) is integer \cite{FisherPRB1989}. 
The system is in the Mott insulator phase when $J/U<(J/U)_c$ and it is in the superfluid phase when 
$J/U>(J/U)_c$. 

The location of the critical point depends on the  dimensionality of the
system and
the filling factor. We are primarily interested here in the two-dimensional
(2D) Bose-Hubbard model  at unit filling factor. Such a model can be emulated in 
a cold atom cloud placed in the optical lattice  generated by three standing 
laser beams producing the periodic potential 
\be
V\cos^2(kx)+V\cos^2(ky)+V_\perp\cos^2(kz), \ \ k=2\pi/\lambda
\label{latte}
\ee
for atoms (see \cite{Porto1,Porto2,Porto3,GreinerSci2010,Endres1,Endres2} for experimental studies
of the 2D Bose-Hubbard model). Above $\lambda$ is the wavelength of the
laser beams, $V$
is the height of the optical potential in the $x$--$y$ plane where we study 
the 2D Bose-Hubbard model, and $V_\perp$ is the height of the 
lattice potential confining the atoms to this plane ($V_\perp\gg V$).
 Additionally, we assume that
atoms are kept in the optical lattice by the optical box trap
enabling studies of homogeneous systems
(see \cite{RaizenPRA2005,HadzibabicPRL2013} for experiments on  cold atoms in box 
potentials).

The position of the critical point in the 2D Bose-Hubbard model at unit
filling factor  has been discussed in numerous theoretical
studies and an agreement has been reached that
\begin{equation}
(J/U)_c\approx 0.06.
\label{critical}
\end{equation}
This value was obtained by theoretical studies of  the superfluid
density \cite{KrauthEPL1991},  compressibility \cite{SorensenPRL2005},
structure factor \cite{CapelloPRB2008}, 
energy gap \cite{MonienPRB1996,AmicoPRL1998,MonienPRB1999,SansonePRA2008},  
entanglement entropy \cite{FrerotPRL2016},  fidelity susceptibility \cite{WangPRX2015},
fixed points of the  real-space renormalization group flow \cite{RokhsarPRB1992}, and 
superfluid order parameter \cite{EckardtPRB2009,SenguptaPRB2012}. 
We will discuss below two more observables that can be used to locate the
critical point. They are in our view the most natural experimentally-relevant 
observables one can think of   in the context of the Bose-Hubbard model.

\section*{Results}
{\bf Idea}. To proceed, it is convenient to introduce the parameter
\be
\xi=J/U
\label{xiii}
\ee
and denote the ground state of the Hamiltonian (\ref{H}) as $|\xi\rangle$. 

Our idea is to locate the critical point of the 2D
Bose-Hubbard model  through the {\it singularities} of the
{\it derivatives} of either the nearest-neighbor two-point correlation function 
\bee
C(\xi)=\left.\la\xi|\hat{a}_{\bf i}^\dag \hat{a}_{\bf j} + 
\hat{a}_{\bf j}^\dag \hat{a}_{\bf i}|\xi\ra\right|_{\la\bf i,j\ra}
\eee
or the variance of on-site atom number operator
\bee
{\rm Var}(\xi)=\la\xi|\hat n^2_i|\xi\ra -\la\xi|\hat n_i|\xi\ra^2,
\eee
which  has
not been done before.

More precisely, we propose that for some specific large enough  $n$
\be
\frac{\partial^n}{\partial\xi^n}C(\xi), \ \ 
\frac{\partial^n}{\partial\xi^n}{\rm Var}(\xi)
\label{sing}
\ee
will be divergent at the critical point in the thermodynamically-large system
at zero absolute temperature.
The singularity of (\ref{sing}) 
follows from the defining feature of a quantum phase transition: 
Non-analyticity of the ground-state energy at the 
quantum critical point (Ch. 1.1 of \cite{Sachdev}; see also
\cite{SachdevToday} for a pleasant  introduction to quantum phase transitions).

We proceed in the standard way to verify this claim.  We 
introduce the ground-state energy ${\cal E} = \la\xi|\hat H|\xi\ra$,
note that ${\cal E}/U$ is a function of $\xi$ only, 
use the Feynman-Hellman theorem 
\be
\frac{\partial}{\partial\xi}\B{\frac{\cal E}{U}}=
\left\la\xi\left|\frac{\partial}{\partial\xi}\B{\hat H/U}\right|\xi\right\ra,
\label{fh}
\ee
and  take into account that 
\be
\la\xi|\sum_{\la\bf i, j\ra}(\hat{a}_{\bf i}^\dag \hat{a}_{\bf j} + 
\hat{a}_{\bf j}^\dag \hat{a}_{\bf i})|\xi\ra
=2MC(\xi)
\label{p1}
\ee
and 
\be
\la\xi|\sum_{\bf i}\hat{n}_{\bf i}(\hat{n}_{\bf i}-1)|\xi\ra=M {\rm Var}(\xi),
\label{p2}
\ee
where $M$ is the number of lattice sites. 
The 2D square lattice geometry and unit filling factor 
are assumed in equations (\ref{p1}) and (\ref{p2}), respectively. 
The translational invariance of the ground state is also assumed 
in these equations.
Generalization of equations (\ref{p1}) and (\ref{p2}) to other lattice
geometries and filling factors is straightforward.

Using equations (\ref{fh}--\ref{p2}) one finds that 
\be
\frac{\partial^{n}}{\partial\xi^{n}} C(\xi) =-\frac{1}{2}
 \frac{\partial^{n+1}}{\partial\xi^{n+1}}\B{\frac{\cal E}{MU}}
\label{dn_1}
\ee
and  
\be
\frac{\partial^{n}}{\partial\xi^{n}} {\rm Var}(\xi) =-
2 \frac{\partial^{n-1}}{\partial\xi^{n-1}}\BB{\xi\frac{\partial^2}{\partial\xi^2}\B{\frac{\cal E}{MU}}}.
\label{dn_2}
\ee

As the quantum critical point is traditionally  associated with  non-analyticity of the
ground-state energy, we assume that the derivatives of the ground-state energy with
respect to the parameter driving the transition, 
$$
\frac{\partial^m}{\partial\xi^m}{\cal E},
$$
are continuous for $m=0,\dots,n$ and
either divergent or discontinuous at the critical point  for $m=n+1$. 

The question now is which derivative of the ground-state energy should we
expect to be divergent? To answer this question, we note that 
the quantum phase transition of the 2D Bose-Hubbard model 
lies in the universality class of the classical 3D XY model \cite{FisherPRB1989},
whose singular part of the free energy scales  with the distance $t$ from the critical
point as $f_s(t)\sim t^{2-\alpha}$, where $\alpha$ equals about $-0.0136$ \cite{VicariPRB2001,HeliumPRB2003}.
Following the discussion in Ch. 1.7 of \cite{Baxter}, we write 
$f_s(t)=f_+(t)-f_-(t)$, where $f_+$ ($f_-$) is the  free energy for $t>0$ ($t<0$),
which has been analytically continued to the  complex $t$ plane.
Non-analyticity at the
critical point is then seen by non-zero derivative(s) of $f_s$ at
$t=0$. The third and higher derivatives of $f_s$ are divergent  at the
critical point of the classical 
3D XY model. Going back from the classical 3D XY model to the quantum 2D
Bose-Hubbard model, we expect that the third derivative of the ground-state
energy will be singular. This translates into the divergent second derivative 
of both  nearest-neighbor correlation function and variance through  (\ref{dn_1}) and (\ref{dn_2}), respectively. 
Such singularity of $C(\xi)$ and ${\rm Var}(\xi)$ is expected to develop in the  limits of temperature $T\to0$ and
the system size $M\to\infty$. For finite systems instead of a singularity 
either  an extremum or a kink smoothing out discontinuity should  develop
near the critical point for small enough temperatures. 
We will now discuss the observables that we use for finding the critical
point.

{\bf Observables.} The correlation functions 
$\la\xi|\hat{a}_{\bf i}^\dag \hat{a}_{\bf j} + \hat{a}_{\bf j}^\dag
\hat{a}_{\bf i}|\xi\ra$
are arguably the most
experimentally accessible  correlation functions in cold atom systems. 
It is so because their Fourier transform provides a quasi-momentum distribution of
cold atom clouds, which can be extracted from  the
time-of-flight images \cite{GreinerNature2002,ProkofevPRA2002}
(see  \cite{Porto1,Porto2,Porto3} for measurements in the 2D Bose-Hubbard
system). 

The critical point can be   extracted from  these correlation
functions through the study of their decay with the distance $|{\bf i}-{\bf j}|$
between the lattice sites. They are expected 
to decay exponentially in the Mott
phase and algebraically in the superfluid phase in the thermodynamically-large zero-temperature system. 
Such a strategy of finding the critical point in the 2D
Bose-Hubbard model is problematic  because the exponential vs. algebraic transition 
is expected to happen for large $|{\bf i}-{\bf j}|$ distances. Such distances are 
hard to deal with  in  theoretical calculations because the model is not exactly
solvable and it does require substantial computational resources to handle 
moderate lattices sizes.
On the cold atom experimental side,  one has to face  issues 
with accurate measurement of distant correlation functions.

Therefore, we would like to argue that our approach 
provides a more practical way of locating the critical point as it
is based on the nearest-neighbor two-point correlation function, which among
other correlation functions is the easiest to obtain both theoretically and experimentally.

\begin{figure}[t]
\includegraphics[width=0.71\textwidth,clip=true]{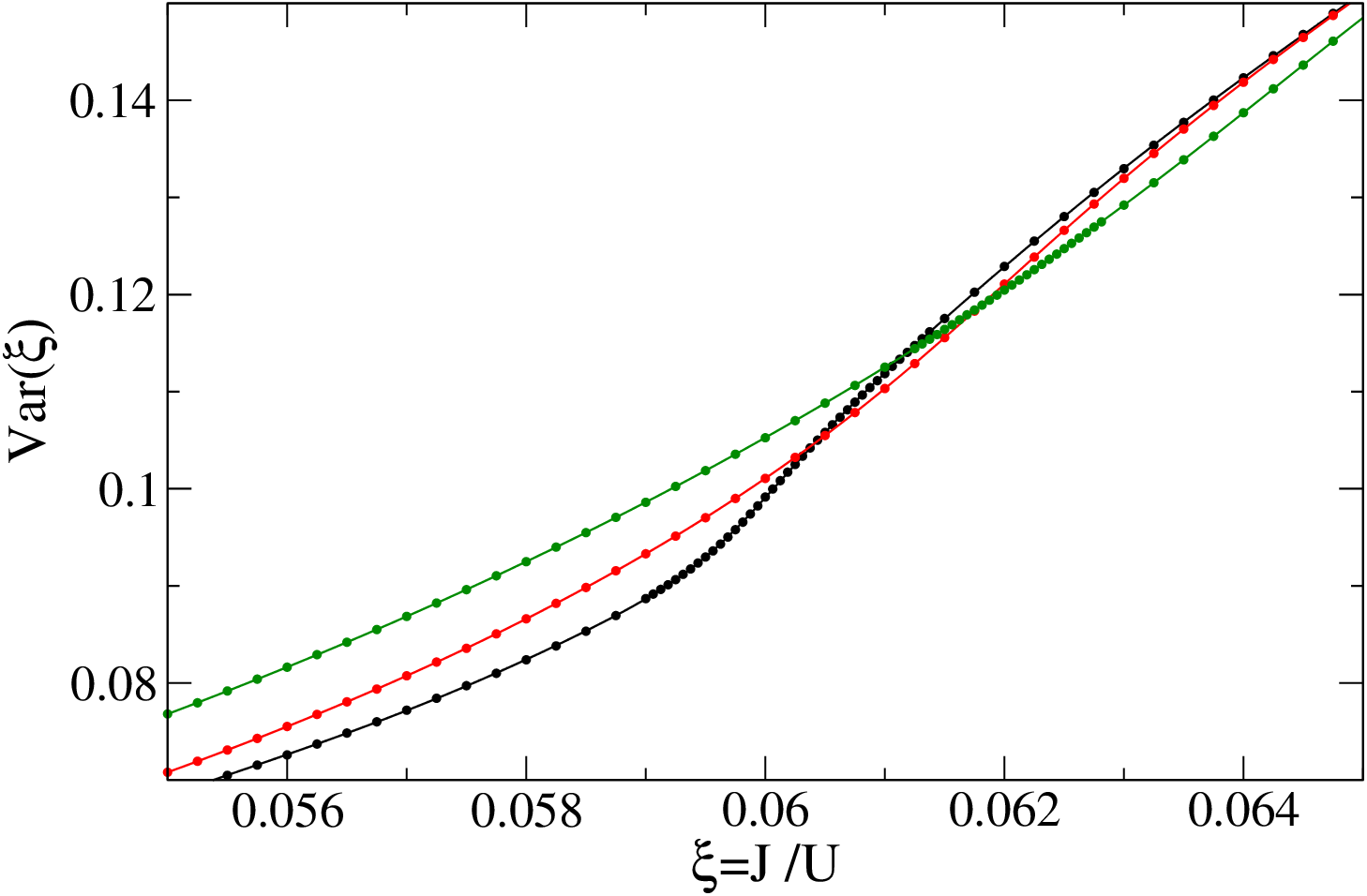}
\caption{The variance of the on-site number operator.
The circles provide QMC data while the solid lines are Pad\'{e}
approximants (\ref{Pade}) fitted to the numerics with parameters $m\le9$ and $n\le6$
(the lower the temperature the higher order polynomials are needed to
fit the numerics).
The curves from bottom to top (in the
left part of the plot) correspond to temperatures $\T$ equal to $0.005$ (black), 
$0.04$ (red), $0.06$ (green), respectively. The system size is  $M=40^2$. 
}
\label{var_fig}
\end{figure}

The variance of the on-site atom occupation can be estimated in-situ thanks to
the recent  breakthrough in the quantum gas microscopy \cite{GreinerPRA2015}.
This technique allows for the detection  of  $0,1,2,3$ atoms in 
individual lattice sites. Choosing the sites far away from the borders of the
trap, one should be able to  minimize the influence of  finite-size
effects, which  should facilitate extraction of the critical point from
the experimental data.

The derivatives of the two observables, $C(\xi)$ and ${\rm Var}(\xi)$,
are proportional to each other \cite{BDNJP2015}
\bee
\frac{\partial}{\partial\xi}{\rm Var(\xi)}= 4\xi\frac{\partial}{\partial\xi} C(\xi),
\eee
so it suffices to measure either one of them.

Therefore, we would like to stress that the measurements of
either $C(\xi)$ or ${\rm Var}(\xi)$ are possible in the current
state-of-the-art experimental setups.  In fact, the measurements 
of $C(\xi)$ have been possible since the seminal paper of Greiner {\it et al} 
\cite{GreinerNature2002}.
It is thus a little bit surprising that nobody has studied 
the derivatives of at least $C(\xi)$ to obtain 
unambiguous signatures of a superfluid-Mott insulator quantum phase transition.
The idea to find the critical point through  the
expectation values (or thermodynamical averages in classical phase
transitions) of  different terms of the Hamiltonian is 
quite natural and has been explored before (see e.g. \cite{Baxter} in the
context of classical and \cite{VenutiPRB2008} in the context of quantum phase transitions). 
To the best of our knowledge, however, {\it non-analytic} properties of these expectation values 
have not been explored in the context of cold atomic systems. We fill this gap by presenting 
the following quantum Monte Carlo simulations 
that we hope will motivate future experimental efforts.

{\bf Quantum Monte Carlo simulations.}
We perform quantum Monte Carlo (QMC)  simulations of the 2D Bose-Hubbard model
(\ref{H}) imposing periodic boundary conditions on the lattice.
We divide the Hamiltonian by $U$, thereby choosing $U$ as the unit of energy,  and set
$\xi$ as the parameter driving  the transition (\ref{xiii}).

\begin{figure}[t]
\includegraphics[width=0.71\textwidth,clip=true]{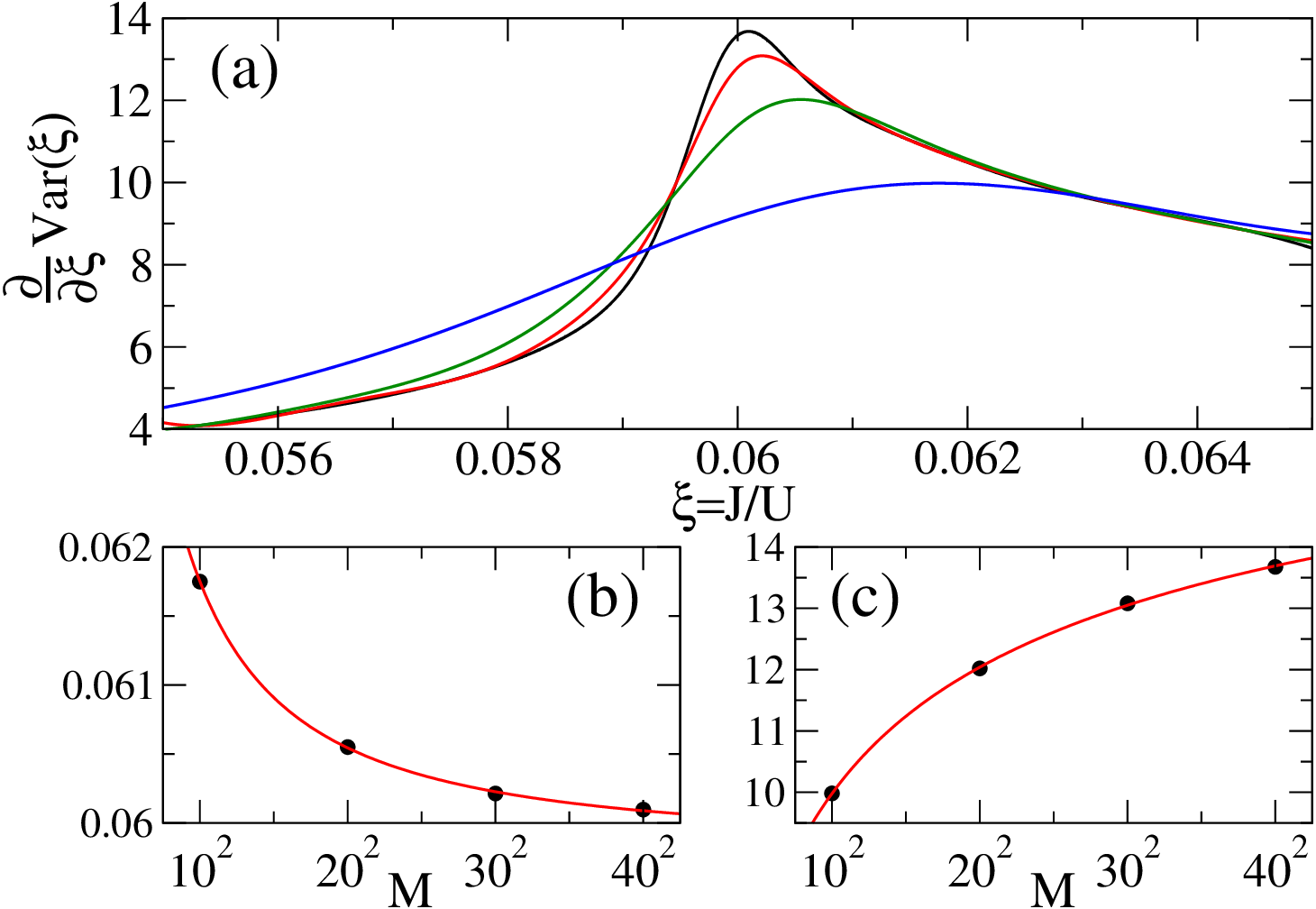}
\caption{
Plot (a): First derivative of the variance of the on-site number operator. 
The curves from top to bottom (around the maximum)
correspond to $M$ equal to $40^2$ (black), $30^2$ (red), $20^2$ (green), and $10^2$
(blue),  respectively.
Plot (b): The position of the maximum of $\pVar$. Plot (c): $\pVar$ at the maximum. 
The circles in plots (b) and (c) show QMC data, while the solid lines provide
 fits (\ref{fit}) and (\ref{fit_}), respectively.
All plots are for $\T=0.01$. 
}
\label{size_fig}
\end{figure}

\begin{figure}[t]
\includegraphics[width=0.71\textwidth,clip=true]{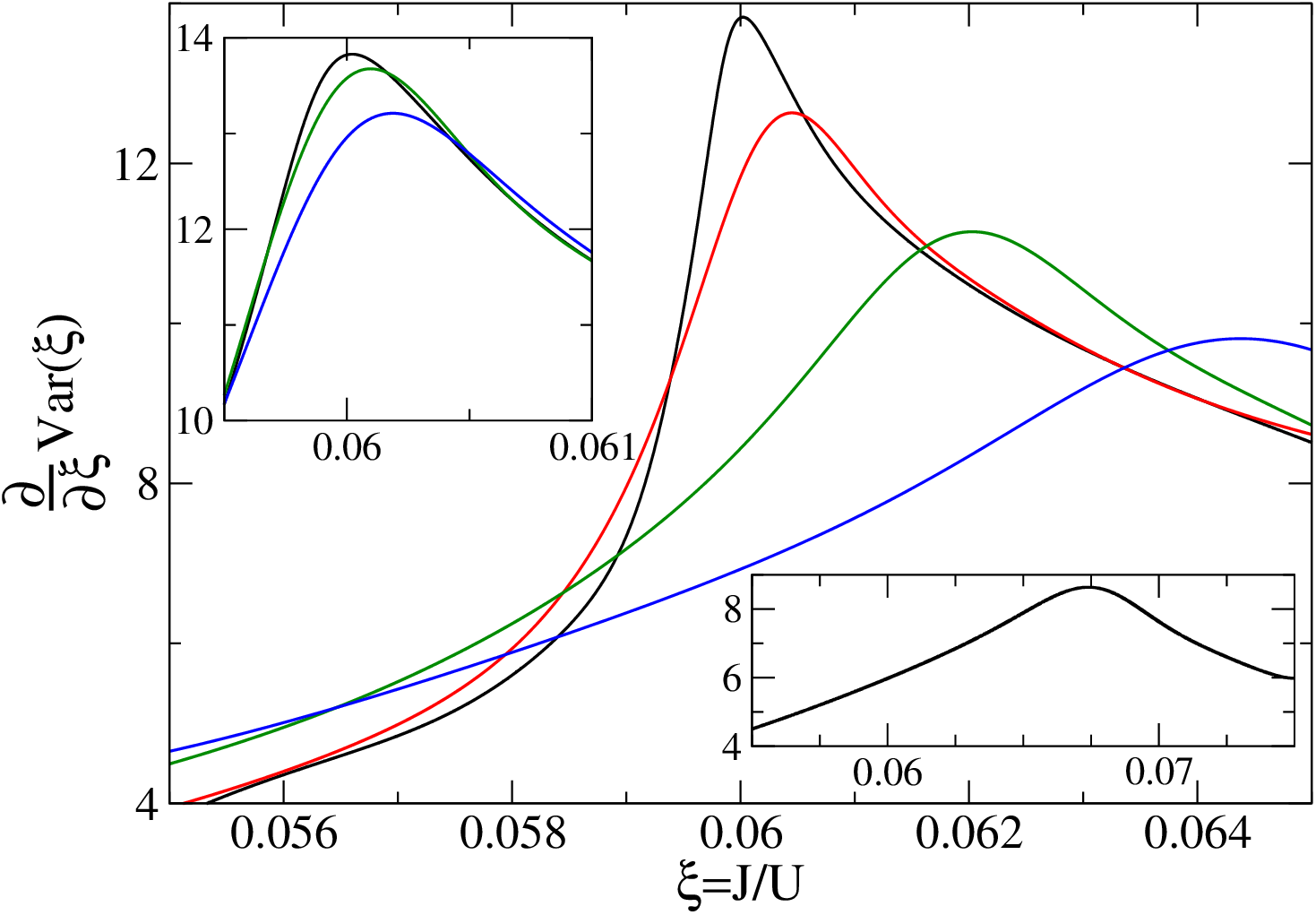}
\caption{Derivative of the variance of the on-site number operator.
The curves from top to  bottom in the main plot 
correspond to temperatures $\T$ equal to $0.005$ (black), $0.02$ (red), $0.04$ (green), 
$0.06$ (blue), respectively. Upper inset illustrates the
convergence of our results towards the $T=0$ limit. It shows our  three
lowest-temperature results: from top to bottom $\T$  equals 
$0.005$ (black), $0.01$ (green), and $0.015$ (blue), respectively. 
Lower inset shows the same as the main plot but for $\T=0.08$.
The system size  is $M=40^2$ for all the calculations in this
figure. 
}
\label{temp_fig}
\end{figure}

\begin{figure}[t]
\includegraphics[width=0.71\textwidth,clip=true]{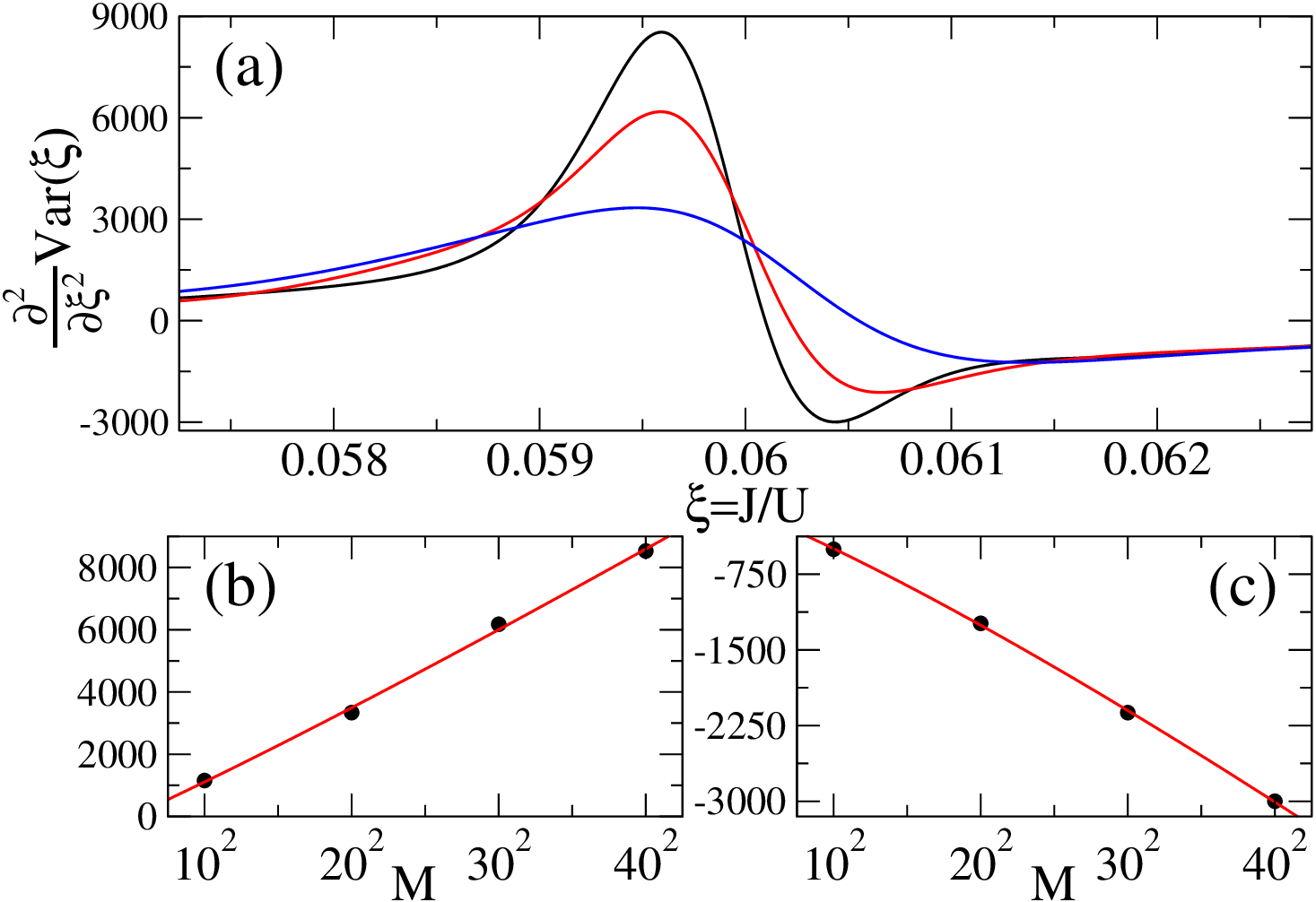}
\caption{
Plot (a): Second derivative of the variance of the on-site number operator. 
The curves from top to bottom (around the maximum)
correspond to $M$ equal to $40^2$ (black), $30^2$ (red), and $20^2$ (blue),
respectively.
Plot (b): $\p2Var$ at the maximum. Plot (c): $\p2Var$ at the minimum. 
The circles in plots (b) and (c) show QMC data, while the solid lines provide
the fit (\ref{fit_}). All plots are for $\T=0.01$. 
}
\label{ddV_fig}
\end{figure}

We use the Directed Worm Algorithm from the ALPS software package
\cite{alps1,alps2}.  To evaluate thermodynamical averages, the algorithm samples the space of
``worldlines'' allowing for the change of the total number of particles. 
To efficiently evaluate the canonical ensemble averages, the
chemical potential is adjusted to yield a unit density in the grand canonical
ensemble, thereby maximizing the total number of samples corresponding to the desired 
average filling factor. In the end, only the ``worldlines'' with the number of atoms  equal to
the number of lattice sites are averaged to yield the results presented in
Figs. \ref{var_fig}--\ref{ddV_fig}.

We compute the variance of the on-site atom occupation for  lattice sizes $M=10^2$ to $40^2$ and 
temperatures $\T$ between $0.005$ and $0.08$. 
To estimate  what such temperatures
correspond to, we assume two plausible experimental setups. Namely,  
$^{23}$Na and $^{87}$Rb atoms placed in the lattice (\ref{latte}) with 
$\lambda=532$nm and $V_\perp=30 E_R$, where the recoil energy $E_R = \hbar^2 k^2/2m$ with $m$ being the
mass of the  atom. As the s-wave scattering lengths we take $2.8$nm
for sodium and $5.3$nm for rubidium. 
Computing  Wannier functions and proceeding in the standard way \cite{BlochRMP2008}, 
we find that the critical point (\ref{critical})  for the sodium (rubidium)  
system is located at the height $V$ of the 
lattice potential equal to $10.1E_R$ ($8.0E_R$), for which 
the coefficient $U$ equals $0.31E_R$ ($0.51E_R$). 
Having these coefficients, one finds that $U/k_B$ at the critical point  equals 
$461$nK for sodium and $199$nK for  rubidium, respectively.
Combining these
results, we see that the highest temperatures that we consider are $37$nK ($16$nK) for 
the above-proposed sodium (rubidium) setup. Both temperatures are
experimentally accessible \cite{Porto3,GreinerSci2010,Endres1,Endres2}. 

The  results that we obtain
are presented in Figs. \ref{var_fig}--\ref{ddV_fig}. To be able to accurately extract the
derivatives of the variance, we fit Pad\'e approximants
\be
{\rm Var}(\xi)=\frac{\sum_{s=0}^m A_s \xi^s}{1+\sum_{s=1}^n B_s\xi^s}, 
\label{Pade}
\ee
to the QMC numerics (Fig. \ref{var_fig}) and then differentiate the resulting curves (Figs. 
\ref{size_fig}--\ref{ddV_fig}). Such a procedure removes the
influence of small fluctuations in the QMC calculations on our results. 
Moreover, it can be straightforwardly applied to experimental
data that will be affected in a similar way by the limited  accuracy of  measurements.

Looking at Fig. \ref{var_fig}, we see that for the lowest temperature displayed, $\T=0.005$, there is 
a steep increase of the variance around the critical point (\ref{critical}). 
Such an abrupt  increase is reminiscent of the  behavior of  magnetization of the 
1D quantum Ising model in the transverse field near the critical
point \cite{PfeutyAnn1970}. As  temperature rises, the abrupt growth of the variance near the
critical point fades away and the variance seems to be featureless, 
which is illustrated for $\T=0.04$ and $0.06$ in Fig. \ref{var_fig}. 
It is thus worth  to stress that the
position of the critical point is  beautifully encoded in all 
the  curves from Fig. \ref{var_fig}.

In order to extract it, we  compute $\pVar$, where
$\partial_\xi=\partial/\partial\xi$,  finding
that it has a maximum very close to the critical point (Fig.
\ref{size_fig}). The position of the maximum,  $\xi_{\rm max}(M)$,  
moves towards the critical point as the system size is increased. 
To extrapolate it to the thermodynamic limit, 
we fit   
\be
\xi_{\rm max}(M)=a + bM^{-c/2}
\label{fit}
\ee
to QMC data for $M=10^2-40^2$ and $\T=0.01$ (Fig. \ref{size_fig}b; all the fits below are also done
for these parameters).    
We obtain $a=0.0598$, $b=0.0491$,  $c=1.40$.
It turns out that the value of the parameter $a=\xi_{\rm max}(\infty)$ is the same
as the  most accurate estimations of the position of the critical point in the
2D Bose-Hubbard model  \cite{MonienPRB1999,SorensenPRL2005,SansonePRA2008}. 

Furthermore, we observe that if we fix the system size and vary 
temperature, then the position of the maximum of $\pVar$  approaches
the critical point when  $T\to0$ (Fig. \ref{temp_fig}).
Moreover,  we observe that $\pVar$  at the maximum grows 
with both the system size (Fig. \ref{size_fig}) and the inverse of the
temperature  (Fig. \ref{temp_fig}).

Therefore, it is interesting to ask whether the studied maximum 
is in fact the singularity that is rounded off and shifted away from the
critical point by finite-size effects.
To investigate it, we fit $\pVar(\xi_{\rm max})$ with
\be
\tilde a +\tilde b M^{\tilde c/2},
\label{fit_}
\ee
getting $\tilde a=20.4$, $\tilde b=-21.7$,  $\tilde c=-0.318$ (Fig.
\ref{size_fig}c). 
Taking the limit of $M\to\infty$, we find that 
instead of a singularity there is a maximum of $\pVar$  in the
thermodynamic limit.

Hunting for a singularity, we compute  $\p2Var$ getting the maximum
and  minimum near the critical point (Fig. \ref{ddV_fig}a). 
The study of $\p2Var$ at the extrema
through the fit (\ref{fit_}) supports the conclusion that there is a singularity appearing in
the thermodynamic limit (Figs. \ref{ddV_fig}b and \ref{ddV_fig}c). Indeed, for maximum (minimum) we get
$\tilde a=-981$, $\tilde b=167$, $\tilde c=1.10$ ($\tilde a=83.2$, 
$\tilde b=-36.0$, $\tilde c=1.21$). This means that as
$M\to\infty$, we have $\p2Var\to\pm\infty$ at the extrema. In-between these 
extrema there is the point where $\p2Var=0$, i.e., where the maximum 
of $\pVar$ is located. Thus, in the thermodynamic limit the 
divergent discontinuity of $\p2Var$ will be located at the same 
point as the maximum of $\pVar$ (if that wouldn't be the case, then there would
be two points where $\p2Var$ is non-analytic, which would contradict presence of a
single critical point in the system). 
This observation explains why the non-singular in the thermodynamic
limit maximum of $\pVar$ encodes the position of the critical point so accurately.

The extrapolated thermodynamic-limit singularity of  $\p2Var$ implies the singularity of the third derivative
of the ground-state energy (\ref{dn_2}), which agrees with the above-presented
discussion based on the scaling theory of phase transitions. It is worth to stress
that it is so because the critical exponent $-1<\alpha<0$. Such an exponent
can be directly experimentally measured near the lambda transition 
in liquid $^4$He, which also belongs to the universality class of the
classical 3D XY
model (see e.g. \cite{HeliumPRB2003} reporting the outcome  of an 
experiment done in a Space Shuttle to eliminate the influence of gravity on the 
 transition). On the theoretical side, one can  obtain this exponent 
through the hyperscaling relation linking it to  more commonly 
studied critical exponents \cite{ContinentinoBook}
\be
\alpha=2-\nu(d+z),
\label{alpha}
\ee
where $\nu$ is the exponent providing
algebraic divergence of the correlation length,  $z$ is the dynamical exponent
relating the excitation gap to the inverse correlation length, and $d$ is the dimensionality of the 
quantum system.
In our system $d=2$ and $z=1$ and so $\alpha\gtrless0$ for $\nu\lessgtr2/3$. Since the
critical exponent $\nu$ is nearly $2/3$ in the 2D Bose-Hubbard model, its
 accurate determination is needed to find out  whether $\alpha$ is a
little bit smaller or greater than zero. Had the latter possibility  been realized,
the second derivative of the ground-state energy (the  first derivative of either 
variance or nearest-neighbor correlation function) would have been divergent.

Finally, we mention that the standard expectation coming from the finite-size scaling theory is that 
thermodynamic-limit singularities are rounded off and 
shifted away from the critical point by the distance $\sim M^{-\chi/d}$,
where $\chi$ is an integer multiple of $1/\nu$  \cite{Cardy1,Cardy2}. 
Fitting the position of the maximum (minimum) of $\p2Var$
 with (\ref{fit}), we obtain $a=0.0596$, $b=-0.542$,  $c=2.74$ ($a=0.0598$, $b=0.0788$, $c=1.30$).
Thus, we see that the fitting parameter $c$ roughly matches  integer multiples of $1/\nu\approx1.49$
in the 2D Bose-Hubbard model \cite{VicariPRB2001,HeliumPRB2003}.
The same conclusion applies to the finite-size scaling of the position of the
maximum of $\pVar$; see the fitting results right below (\ref{fit}).
Simulations of larger system sizes are needed for making conclusive predictions about the
relation between $c$ and $1/\nu$.

\section*{Discussion}
\label{Sum}

We have shown  that  derivatives of the 
nearest-neighbor correlation function and the variance of on-site atom number 
operator can be used as an efficient and 
experimentally-relevant  probe of the location of the critical point of the 
2D Bose-Hubbard model.

Similar calculations can be performed for the 
3D Bose-Hubbard model, where $z=1$ and  $\nu=1/2$
\cite{FisherPRB1989}. Using equation (\ref{alpha}) one then finds that 
$\alpha=0$. Therefore, based on the discussion from the Results section, we
expect that the second derivative of the ground-state energy as well as the  
first derivative of nearest-neighbor correlation function and the variance of
the on-site atom number operator will be  divergent at the critical point of this model. 
This conclusion applies to the thermodynamically-large zero-temperature 
system, while in the finite-size system we expect to find an extremum of these
observables near the critical point just as in the 2D Bose-Hubbard model.

Interestingly, in the 1D  quantum Ising model in the transverse field, where
$z=\nu=1$, $\alpha=0$ as well. In this model a closed-form expression for the ground-state energy 
is known \cite{PfeutyAnn1970} and one can  check that indeed the second derivative of the 
ground-state energy is divergent at the critical point of this model.

Furthermore, we would like to mention that it is unclear at the moment 
whether one can extract the position of the critical point of the 1D
Bose-Hubbard model in a similar way. The problem is that such a model 
undergoes a Berezinskii-Kosterlitz-Thouless  (BKT) transition \cite{FisherPRB1989}, 
where the singularities 
associated with the critical point are not algebraic but exponential in the distance from the
critical point \cite{PanditPRL1996}. 
As a result, the above discussion of the singularity of the ground-state energy 
based on the exponent $\alpha$ is not readily applicable  as it 
assumes that the divergence of the correlation length  is algebraic $\sim |t|^{-\nu}$,
where again $t$ is the distance from the critical point.

Finally, we would like to mention  that the discussion from the Results
section can be straightforwardly extended to Hubbard-like models undergoing a regular,
i.e. not a BKT-type,  transition.  Different such models can be studied with cold
atoms in optical lattices and  one can consider not only bosonic but also  fermionic systems
\cite{Lewenstein,KubaReview}. For example, suppose that  the Hamiltonian of some Hubbard-like model 
 contains the 
nearest-neighbor tunnelling term such as (\ref{H}). 
One can argue then that some derivative of the nearest-neighbor correlation
function with respect to the tunnelling parameter $J$ should be divergent if the change 
of $J$ induces a quantum phase transition. Such a statement should be correct 
regardless of the specific form of the interaction part of the Hamiltonian,
which  can contain other than on-site terms. 
This observation should be useful  in both theoretical and experimental studies of the location of the 
critical point in ubiquitous Hubbard-like models. \\

\noindent{\bf Acknowledgments}

\noindent M\L\  was  supported by Foundation for Polish Science (FNP) and the ERC
Synergy Grant UQUAM. BD was supported by the  Polish National Science Centre  project
DEC-2013/09/B/ST3/00239. JZ was supported by the Polish National Science
Centre  project DEC-2015/19/B/ST2/01028. Partial  support by
 PL-Grid Infrastructure and EU via project
QUIC (H2020-FETPROACT-2014 No. 641122) is also acknowledged.\\


\end{document}